\documentclass[manuscript,screen]{acmart}
\usepackage{subcaption}

\AtBeginDocument{%
  }

\setcopyright{acmlicensed}
\copyrightyear{2018}
\acmYear{2018}
\acmDOI{XXXXXXX.XXXXXXX}
\acmConference[Conference acronym 'XX]{Make sure to enter the correct
  conference title from your rights confirmation email}{June 03--05,
  2018}{Woodstock, NY}
\acmISBN{978-1-4503-XXXX-X/2018/06}

\begin{document}

\title[Reflection in Human--LLM Co-Creation]{Pinning ``Reflection'' on the Agenda: Investigating Reflection in Human--LLM Co-Creation for Creative Coding}



\author{Anqi Wang}
    \authornote{Authors are affiliated with HKUST, Hong Kong SAR.}
    \authornote{Authors are affiliated with HKUSTGZ, China.}
\author{Zhizhuo Yin}
    \authornotemark[2]
\author{Yulu Hu}
    \authornotemark[2]
\author{Yuanyuan Mao}
    \authornotemark[2]
\author{Lei Han}
    \authornotemark[2]
\author{Xin Tong}
    \authornotemark[2]
\author{Keqin Jiao}
        \authornote{Author is affiliated with Carnegie Mellon University, United States.}
\author{Pan Hui}
 \authornotemark[2]\authornotemark[1] 
    \authornote{Corresponding author.}



\renewcommand{\shortauthors}{Wang et al.}

\begin{abstract}

Large language models (LLMs) are increasingly integrated into creative coding, yet how users reflect, and how different co-creation conditions influence reflective behavior, remains underexplored. This study investigates situated, moment-to-moment reflection in creative coding under two prompting strategies: the entire task invocation (T1) and decomposed subtask invocation (T2), to examine their effects on reflective behavior. Our mixed-method results reveal three distinct reflection types and show that T2 encourages more frequent, strategic, and generative reflection, fostering diagnostic reasoning and goal redefinition. These findings offer insights into how LLM-based tools foster deeper creative engagement through structured, behaviorally grounded reflection support. 
\end{abstract}


\received{20 February 2007}
\received[revised]{12 March 2009}
\received[accepted]{5 June 2009}

\makeatletter
\def\@authorsaddresses{}
\def\@setauthorsaddresses{}
\makeatother


\maketitle

\section{Introduction}

Recent advances in large language models (LLMs), such as Code Llama and ChatGPT-4, which demonstrated significant capabilities in interpreting natural language descriptions of programming tasks~\cite{hamalainen2023evaluating, bang2023multitask, liu2023evaluating} and generating functional, contextually appropriate code~\cite{chakraborty2022natgen, dakhel2023github, fried2022incoder, nijkamp2022conversational, castelvecchi2022chatgpt}, has been applied to broad art and design domains~\cite{wang_survey_2025}. These developments have enabled a novel form of \textbf{human–AI co-creation} in \textbf{creative coding} (e.g.,~\cite{angert2023spellburst})—one that combines LLMs’ technical fluency with the iterative, exploratory nature of artistic practice. 
Creative coding~\cite{peppler2005creative} emphasizes artistic conceptsand visual appeal through iterative codes, yet it presents unique challenges: artists must both imagine system-level behaviors and operationalize them through complex programming environments like P5.js~\cite{mccarthy2015getting}. 
\textbf{\textit{Reflection}} is central to navigating this creative complexity, enabling artists to evaluate goals, reframe strategies, and evolve thinking~\cite{guillaumier_reflection_2016,ford_role_nodate,candy_creative_2020}.
Moreover, it fosters self-awareness, allowing users to adapt thinking and overcome obstacles, optimizing creative solutions.
In human-LLM co-creation, reflection takes on new dynamics: users respond not only to their work but also to machine-generated suggestions, bugs, or alternatives. This shifts reflection from purely internal evaluation to an interactive process shaped by dialogic rhythm and prompt framing. The mode of co-creation—whether through whole-program generation or decomposed subtasks—can influence when and how reflection occurs, and whether it leads to surface-level fixes or deeper conceptual reframing. 


    However, existing research largely overlooks how reflection unfolds within human-LLM co-creation, focusing instead on technical performance or post-hoc interpretation. To our knowledge, only \citet{ford_reflection_2024} has investigated AI-supported reflection in the context of music composition. What remains unclear is how different forms of reflection emerge in real-time during co-creative dialogue with LLMs, and how prompt framing might shape that process. 
    To address these issues, we investigate reflection as a situated, behavioral phenomenon during creative coding with LLMs under different co-creation approaches. Our study asks: \textit{How do artists reflect during creative coding with LLMs, and how do different types of reflection shape the human–LLM co-creation process under varying co-creation conditions?}
    We conducted our study in two co-creation ways with LLMs involving art practitioners (N=22) with diverse programming backgrounds, one invoking the entire program (T1), the other subtasks solving a problem (T2).  Through qualitative coding and frequency analysis, 
    we found that the T2 condition increased reflection frequency and shifted it from reactive repair to proactive goal reframing and strategic reasoning, highlighting the role of such strategy in fostering deeper reflection.
    This contributes: (1) how Baumer’s reflection typology applies and adapts in the context of LLM-based creative coding, revealing distinct patterns—in the moment-to-moment flow; 
    (2) Empirical evidence  showing how task-decomposed prompting fosters deeper, more strategic reflection; 
    (3) Design implications for rethinking reflective prompting beyond productivity scaffolding, but as a catalyst for creative reasoning and adaptive goal-setting. 

\section{Related Works}
Recent advances in Large Language Models (LLMs) like ChatGPT-4 and Code Llama have enabled them to understand coding tasks from natural language prompts~\cite{hamalainen2023evaluating, bang2023multitask, liu2023evaluating} and produce coherent, contextually appropriate code~\cite{chakraborty2022natgen, dakhel2023github, fried2022incoder, nijkamp2022conversational, castelvecchi2022chatgpt}. These breakthroughs have spurred interest in using LLMs as co-creators in creative tasks (e.g.,~\cite{angert2023spellburst, chen2023large, swanson2021story, bhavya2023cam, di2022idea}, including tools that support creative coding. For example, \citet{angert2023spellburst} developed an environment enabling artists to generate and refine generative art through interaction and direct code edits. However, much of this work emphasizes boosting productivity via LLMs, often neglecting the artist’s reflective role, potentially leading to over-reliance and diminished creative agency in the final outcome.

Reflection is a foundational concept in HCI~\cite{sengers2005reflective}, particularly in creative domains like writing, design, and project documentation~\cite{dalsgaard2012supporting,hansen2012productive,dijk2011noot,hailpern2007team,hummels2009reflective,mendels2011freed,sharmin2013reflectionspace,yoo2013value, dalsgaard2012reflective,sterman2023reflective}. It includes deliberate thought for problem understanding~\cite{sedig2001role} and the re-evaluation of past experience~\cite{bateman2012search}. Baumer et al. offered a comprehensive reflection framework~\cite{baumer_reviewing_2014, baumer_reflective_2015}, identifying three components:
\textbf{Breakdown}: disruptions that prompt reevaluation;
\textbf{Inquiry}: data collection and conceptual exploration;
\textbf{Transformation}: shifts in understanding or creative direction.
Despite its relevance, empirical work on reflection in LLM-assisted creativity remains scarce, motivating our study of how reflection unfolds in creative coding with LLMs.

\section{Methods}
To investigate reflection in human–LLM co-creation on creative tasks, we conducted a mixed-method study with controlled experiments and qualitative interviews. Building on prior design guidelines for open-ended objectives~\cite{ford_reflection_2024}, the study received ethics approval, with all participants providing signed consent and voluntarily participating without compensation. We used the Code-Llama-Instruct-34B model~\cite{roziere2023code} for its strong code generation and natural language capabilities. As Meta released only model weights, we built a custom ChatGPT-style frontend (Appendix Figure~\ref{fig:screenshot}) and deployed the model on an Ubuntu server with 6×NVIDIA A800 80GB GPUs.



\textit{Procedure.} 
Through e-application forms collecting demographic information, posted online in social media groups and offline campus in mainland of China, we screened and recruited a total of 22 participants attended our study with balanced backgrounds (see Appendix Table~\ref{tab:participants}). Our participants are artist- and design-researchers who have wide creative coding needs in their daily creation. 
We conducted face-to-face studies in a campus laboratory, with each participant attending individually. 
Following prior studies on human-LLM co-creation paradigms~\cite{chen2023large, ross2023programmer, jonsson_cracking_2022}, we conducted a within-subjects design where each participant completed both conditions (T1 and T2)-two co-creation ways~\cite{chen2023large, ross2023programmer,jonsson_cracking_2022}): \textbf{Condition T1}-invoking Llama with the entire coding goal and \textbf{Condition T2}-decomposing the programming task into subtasks and sequentially invoking the LLM to address each, by collecting observation, think-aloud, and interview method (See Appendix~\ref{appendix:interview}). 
The order of conditions was counter-balanced across participants to control for sequence effects. 
In both two conditions, we asked participants to create a creative coding in the same, self-defined themes but in different forms using \textit{P5.js}~\footnote{\textit{P5.js} is a JavaScript library designed for creative coding on web browsers, facilitating the creation of interactive visuals through code} web-end under two conditions. Required guidelines includes 1) with a minimum two visual elements 2) in interactive and 3) visual-appeal way in each task. 
Before each condition, we explained the prompting strategy with examples, and verified understanding via a brief quiz. We analyzed possible sequence effects by comparing reflection frequency and type between orders and found no significant order effect. 
The study procedure is shown in the Appendix Figure~\ref{fig:studyflow}.

\textit{Data Analysis.}
To analyze reflection behaviors, we employed a mixed-method approach combining qualitative and quantitative analysis. 
We began by conducting a theory-informed thematic analysis, drawing on Baumer’s reflection framework~\cite{baumer_reviewing_2014, baumer_reflective_2015} to identify three primary reflection types: Breakdown, Inquiry, and Transformation. These categories were refined through an iterative process during data collection, and reflection was defined as any user behavior involving rethinking, questioning, or revising ideas in response to LLM feedback.
Both deductive and inductive strategies were applied: the deductive approach mapped behaviors to Baumer’s framework, while the inductive approach captured emergent behaviors beyond the system, such as independent code editing, logic reformulation, or drawing on prior experience.
We segmented the data into reflection episodes, defined as a logic unit comprising a user prompt, LLM output, a reflection trigger (e.g., error or surprise), the user’s response, and the resulting course of action. Each episode was annotated with causal and conditional links (e.g., ``LLM misunderstanding → user clarifies syntax'', ``unexpected output → user shifts goal'') using axial coding principles~\cite{vollstedt2019introduction}.
A codebook was iteratively developed by two researchers, beginning with 5 transcripts coded jointly to establish shared understanding. After alignment on category definitions and coding granularity, the researchers independently coded all reflection episodes. Coder training included discussions on ambiguous cases and guided application of Baumer's definitions. Inter-rater reliability was assessed using Cohen’s $\kappa = 0.84$, indicating strong agreement. Disagreements were resolved through in-depth discussions until full consensus was reached. This process yielded a refined, behaviorally grounded coding scheme that supported both qualitative insights and statistical comparison.
For the quantitative analysis, we analyzed the frequency of reflection instances using a non-parametric Aligned Rank Transform (ART) ANOVA~\cite{10.1145/1978942.1978963}, due to the non-normal, count-based nature of the data. The Reflection Category (Breakdown, Inquiry, Transformation) and Condition (T1, T2) were treated as fixed factors, and the reflection frequency served as the dependent variable. After assessing main and interaction effects, we conducted post-hoc pairwise comparisons and multi-factor contrasts using the ART-C procedure\cite{10.1145/3472749.3474784} to further examine differences between reflection types and conditions.

\begin{table*}[t]
\centering
\scriptsize
\scriptsize\caption{
Three reflection types characterize behaviors in: (1) interactive dialogues with LLMs; and (2) creative coding practices.
}
\begin{tabular}{p{0.2cm}p{1cm}p{6cm}p{6.5cm}}
\textbf{NO.} & \textbf{Category} & \textbf{Dialogues with LLMs} & \textbf{Other Behaviors} \\
\toprule
1 & Breakdown & 1) refining prompts in LLMs for precise requirement descriptions, \newline 2) triggering debug in LLMs through copy-paste error reminders, \newline 3) seeking explanations from LLMs for programs or codes. & 1) user-initiated debugging without LLM co-creation, \newline 2) searching and analyzing materials/documents for error explanations and debugging, \newline 3) referencing provided sources by copying and pasting programs or codes. \\
\hline
2 & Inquiry & 1) optimizing output results based on the current program,\newline 2) seeking information or requesting programming assistance within the design process. & 1) modifying programming or codes to present results accurately based on prior knowledge. \\
\hline
3 & Transfor-\newline mation & 1) adjusting the initial idea or project due to unsuccessful results, \newline 2) responding to unexpected results, \newline 3) proposing an updated problem-solving structure to LLMs. & 1) drawing inspiration or generating ideas from programs provided by external sources. \\
\label{tab:reflectiontypes}
\end{tabular}
\end{table*}

\section{Findings}


    


\paragraph{Types and Frequencies of Reflection}


 

\begin{figure}[t]
    \centering
    \begin{subfigure}[b]{0.28\textwidth}  
        \centering
        \includegraphics[width=\linewidth]{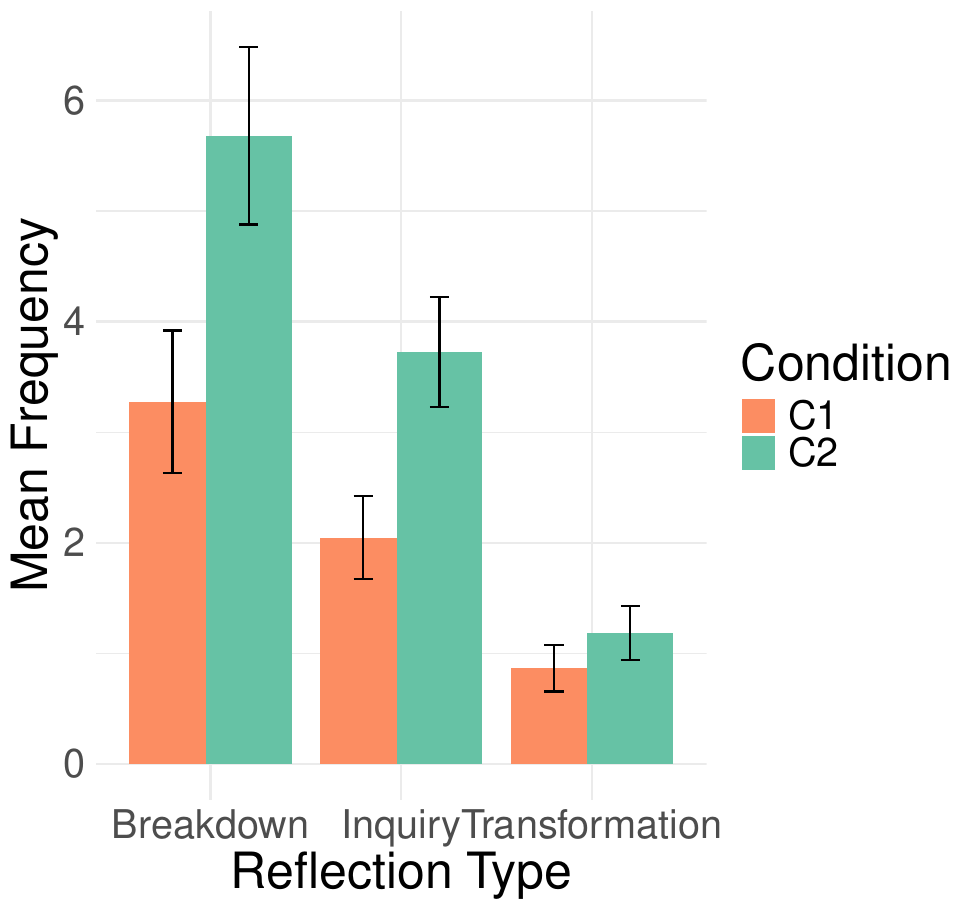}
        \caption{}
        \label{fig:subfig1}
    \end{subfigure}
    \hfill  
    \begin{subfigure}[b]{0.28\textwidth}  
        \centering
        \includegraphics[width=\linewidth]{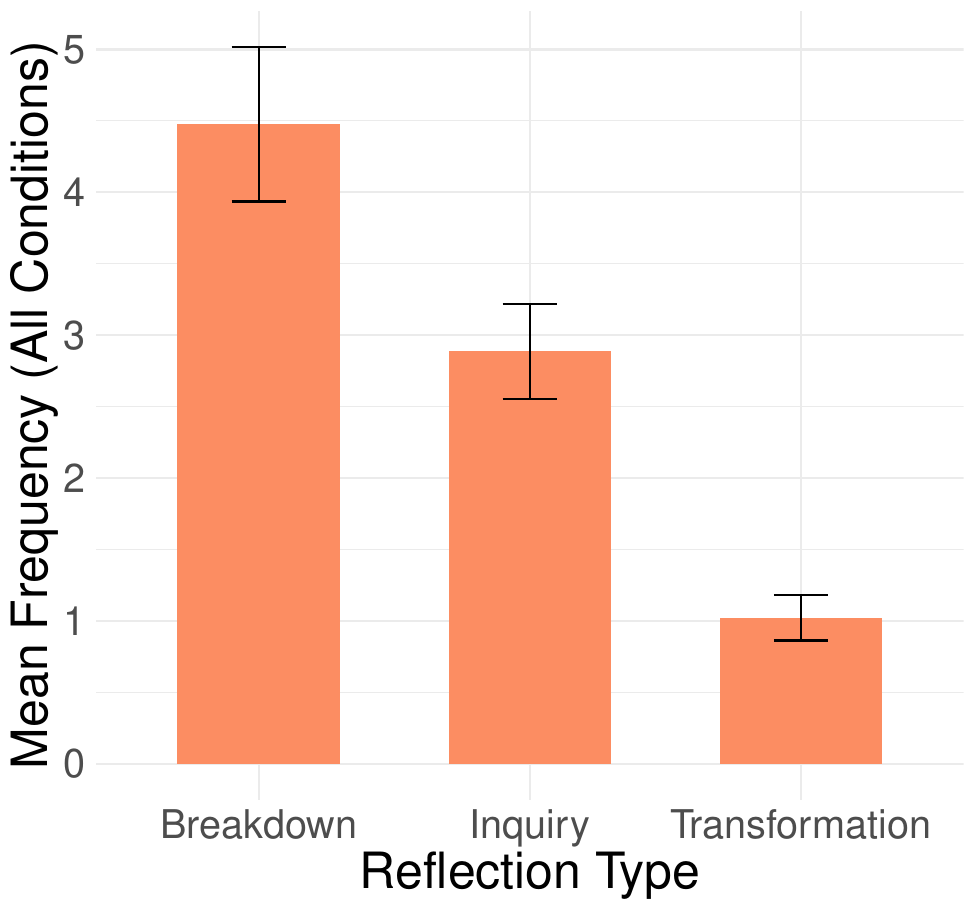}
        \caption{}
        \label{fig:subfig2}
    \end{subfigure}
    \hfill  
    \begin{subfigure}[b]{0.28\textwidth}  
        \centering
        \includegraphics[width=\linewidth]{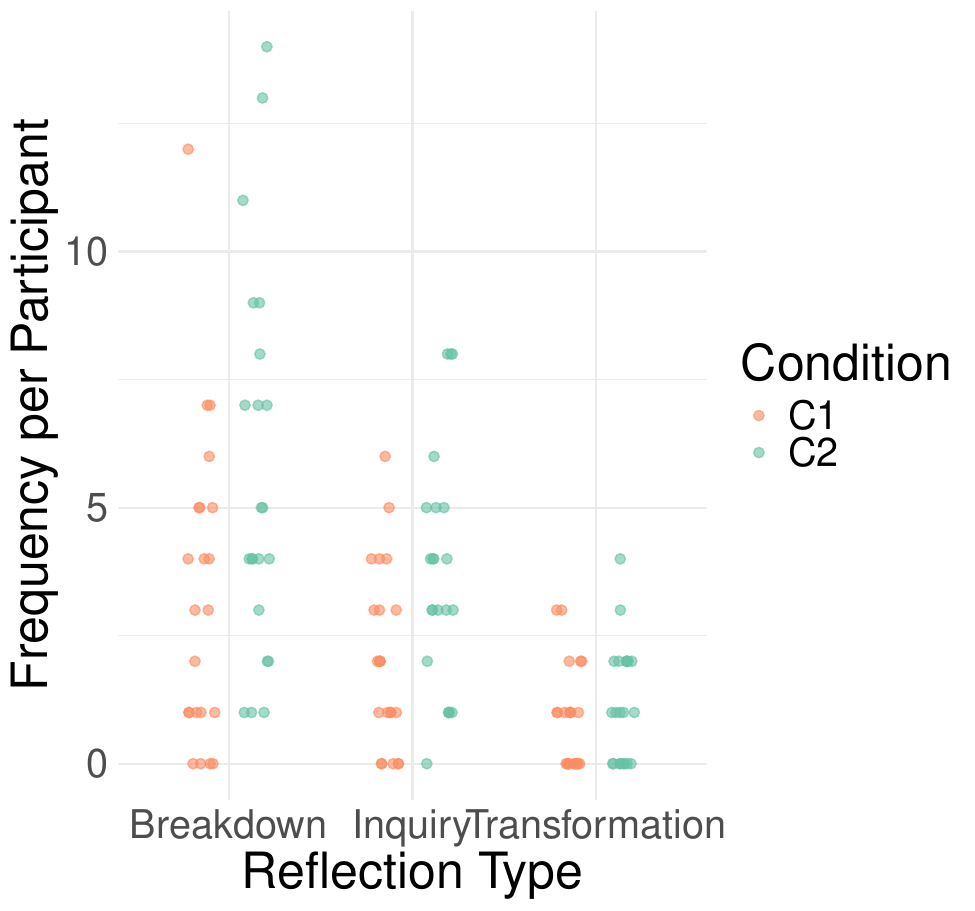}
        \caption{}
        \label{fig:subfig4}
    \end{subfigure}
    \caption{(a) The frequency of reflection types at T1 and T2, with mean values and error bars. 
(b) The overall frequency of each reflection type. 
(c) Individual participants’ reflection type frequencies in T1 and T2.}
    \label{fig:data}
\end{figure}

We identified three types of reflection based on user dialogues with LLMs and parallel behaviors in creative programming as Table~\ref{tab:reflectiontypes}: \textbf{Breakdown}: Triggered by errors or misunderstandings. Users refined prompts, copied errors into LLMs for debugging, or asked for explanations. Outside the LLM, they debugged independently, searched for explanations, and reused code from external sources. \textbf{Inquiry}: Emerged when users sought better outcomes or clarification. Dialogues focused on output refinement or programming help. In non-LLM behaviors, users applied prior knowledge to revise code and improve functionality. \textbf{Transformation}: Occurred when users revised their project direction due to failure or surprise. They reformulated goals in LLM interactions and adapted ideas from external inspirations outside the system. These types reflect distinct reflective strategies across the coding process, from clarifying errors to evolving creative intent.


Descriptive statistics indicated that Breakdown reflections were the most frequent (total $N$ = 203), followed by Inquiry ($N$ = 123), and Transformation ($N$ = 43). 
Compared two conditions, participants reflected more frequently in T2 ($N = 222$) than in T1 ($N = 147$), with a breakdown of 118 Breakdowns, 80 Inquiries, and 24 Transformations in T2, versus 85 Breakdowns, 43 Inquiries, and 19 Transformations in T1. Across conditions, \textbf{Breakdown reflections were most prevalent}, followed by Inquiry and then Transformation.

Our non-parametric Aligned Rank Transform (ART) ANOVA statistical analysis~\cite{10.1145/1978942.1978963} showed a significant main effect of co-creation condition on reflection frequency, ($F(1, 105) = 16.02$, $p < .001$), highlighting the variability in reflection frequencies across different types of reflection (Breakdown, Inquiry, and Transformation)(Figure~\ref{fig:subfig1}). The result also confirmed a significant main effect of Reflection Category ($F(2, 120) = 13.90$, $p < .0001$) (Figure~\ref{fig:subfig2}), indicating that different reflection types occurred with significantly different frequencies overall (Figure~\ref{fig:subfig2}). This suggests that the type of reflection is influenced by both the co-creation condition and participants' reflection category. 
While the interaction between condition and reflection category was not statistically significant ($F(2, 105) = 2.76$, $p = .068$), the effect size was moderate ($\eta^2 = .05$). 
This suggests that, despite limited power to detect a significant interaction, there may be a meaningful behavioral difference in how different types of reflection are shaped by the cooperation condition. 


To probe these patterns, we conducted post-hoc pairwise and multi-factor contrast tests following the ART-C procedure~\cite{10.1145/3472749.3474784}. 
Post-hoc pairwise comparisons showed that \textbf{Breakdown} reflections occurred significantly more often than \textbf{Inquiry} ($p$ < .01), and both occurred significantly more frequently than \textbf{Transformation} ($p$ < .001). 
Further contrast tests revealed that the \textbf{T2} condition led to significantly more Breakdown reflections (T2: $\mu$ = 5.68, T1: $\mu$ = 3.27, $p$ < .01) and Inquiry reflections (T2: $\mu$ = 3.72, T1: $\mu$ = 2.04, $p$ < .05) than the \textbf{T1} condition. Although Transformation reflections were also more frequent in T2 ($\mu$ = 1.18) than in T1 ($\mu$ = 0.86), this difference did not reach statistical significance ($p$ = .37). 
These results suggest that the more scaffolded T2 condition encouraged greater engagement in both surface-level (Breakdown) and diagnostic (Inquiry) forms of reflection, while the deeper, goal-shifting Transformation reflections remained relatively infrequent across both conditions.
Figure~\ref{fig:subfig4} presents a jitter plot of individual participants' reflection type frequencies in T1 and T2, highlighting the distribution of Breakdown, Inquiry, and Transformation reflections.

\paragraph{Reflection Usages}
\textbf{T1: Expressive Ambiguity and Affective Struggle.} In T1, reflection is often triggered by a \textit{Breakdown} in communication between the user and the system, leading to frustration and confusion. Users struggle with ambiguous or overly complex goals, frequently expressing dissatisfaction or emotional fatigue as they attempt to work through these challenges. For example, when encountering a technical failure, users may question their own prompt clarity, as seen in the case of P7, who remarked: \textit{``Maybe I wasn't clear enough… I thought it would understand what I meant.''} In this stage, users tend to focus on surface-level fixes, such as rephrasing prompts or inspecting syntax, without fundamentally addressing the underlying logic or system limitations. 
As P11 demonstrated through repeated phrasing adjustments without questioning the task's conceptual framework—showing minimal progress—they stated: \textit{``I've rewritten this five times, and the output remains unchanged.''} When faced with persistent failure, T1 users often simplify or abandon their original goals, opting for less complex tasks in response to accumulated frustration. For instance, P8 reflected after repeated \textit{Breakdowns}: \textit{``This interaction is too hard—I’ll just make a static visual.''} This response highlights the tendency to scale back ambitions instead of reframing goals in a way that allows for creative expression within the system’s limitations. Despite this, T1 users occasionally re-engage with the system after resetting their goals, though they remain cautious, as demonstrated by P3: \textit{``Let’s see if I can bypass the old problem this time.''} Overall, T1 reflection can be described as emotion-driven, syntactically oriented, and prone to stagnation in the face of system unpredictability or misalignment. 

\textbf{T2: Strategic Adaptation and System-Aware Reasoning.} In contrast, T2 users adopt a more strategic and system-aware approach to reflection, characterized by \textit{Inquiry} and iterative exploration. After an initial \textit{Breakdown}, these users engage in diagnostic reasoning and proceed with a step-by-step exploration of potential solutions, continually refining their approach based on system feedback. For example, P5 demonstrated this by adjusting a parameter to fine-tune the system's output: \textit{``I found that speed 0.5 is too slow, let me try 1.''} Such users prioritize precise control over variables and actively test hypotheses to achieve desired outcomes, with each failure offering insight into further adjustments. In the case of P2, the user identified a conflict in the system’s behavior and proactively modified the prompt’s logic: \textit{``Maybe the delay command conflicts here—I’ll try replacing it with frameCount.''} Rather than abandoning their goals after failure, T2 users tend to reframe or \textit{Transform} their original goals while preserving the underlying creative intent. This is evident when P9 reflected on how to modify their approach to maintain fluidity in the system, saying: \textit{``What I wanted was a sense of fluidity—maybe I can use particles instead of a path.''} Such responses highlight the metacognitive shift that occurs, as users adapt their strategies to better align with what is feasible within the system. T2 users also exhibit the ability to adjust their strategies dynamically, as demonstrated by P6, who diagnosed a missed detail and made a targeted correction: \textit{``I changed the parameter but didn’t reset the system state—maybe that’s the issue.''} Finally, T2 users demonstrate compressed metacognition, rapidly drawing from accumulated knowledge and real-time feedback to make agile decisions. P10, for example, used past experience to quickly identify a potential issue with the canvas, stating: \textit{``I’ve seen canvas issues before—this time it’s probably the loop interfering.''} This ability to synthesize prior learnings and adjust in real-time enables sustained momentum through the creative process, even when faced with setbacks.

\section{Discussion}

Our analysis revealed a consistent pattern across both quantitative and qualitative data: while Condition 2 (T2) increased the frequency of reflection, it also shifted its nature—from reactive responses in T1 to more strategic, generative forms in T2. Although the interaction effect in the ANOVA was marginally significant ($p = .068$), qualitative evidence clearly supported this trend. In T1, reflection was often reactive and triggered by breakdowns, leading users to clarify or simplify goals. In contrast, T2 encouraged more deliberate, strategy-oriented reflection—users manipulated parameters purposefully, compared alternatives, and reframed goals. These behaviors suggest that structured, decomposed prompts in T2 facilitated diagnostic reasoning and evaluative reflection, especially during ideation and transformation. 

These findings extend prior research on AI-assisted creativity~\cite{ford_reflection_2024}, which has primarily emphasized high-level conceptual reflection or post-hoc interpretation. Our work advances this perspective by introducing a \textbf{fine-grained, behaviorally grounded typology of reflection—Breakdown, Inquiry, and Transformation—that emerges in the moment-to-moment flow of LLM-mediated creative programming}. This shift from retrospective abstraction to situated, real-time reflection offers a new lens for understanding reflective engagement as it unfolds dynamically within human–AI co-creation. 
Beyond theoretical insight, our findings carry important implications for co-creative system design. Existing prompt tools largely emphasize productivity—generating correct, complete, or efficient outputs~\cite{mcnutt_study_2023,angert2023spellburst,denny2023promptly,jiang_genline_2021,kazemitabaar_codeaid_2024,zhou_instructpipe_2025}. However, our study reveals a complementary role for prompts: \textbf{provoking strategic reflection and enabling shifts in user reasoning}. To support this, systems should \textbf{adopt cooperative prompting strategies that introduce mild uncertainty, unfamiliar reasoning paths, or contrasting perspectives}. Interface modes might prioritize exploration over immediate solution-finding. For HCI practitioners, this suggests moving beyond success-oriented scaffolding to also deliberately surface productive struggle—guiding users to examine the “why” and “how” of their process at moments that matter most. 

\section{Limitation and Conclusion}
While our study reveals how reflection unfolds in situated, short-burst interactions, future work should examine whether our typology—Breakdown, Inquiry, Transformation—applies to other creative domains such as writing or design. Further, to support broader adoption, systems should adapt cooperative prompting strategies to different user profiles and dynamically surface reflection-supportive modes based on real-time interaction cues. 
Despite these, this work offers a reflection-centered approach for LLM-assisted creativity in HCI. By distinguishing reflection types across two co-creation approaches, we show how LLM configurations influence reflection's frequency and nature. These insights provide design implications for tools that support deeper cognitive engagement beyond task completion.

\begin{acks}
This work was supported in part by the National Key Research and Development Program of China under Grant 2024YFC3307602, the Guangzhou Municipal Nansha District Science and Technology Bureau under Contract No.2022ZD01, and the Guangdong Provincial Talent Program, Grant No.2023JC10X009. 
\end{acks}

\bibliographystyle{ACM-Reference-Format}
\bibliography{sample-manuscript,references}

\appendix
\section{Supplementary Figures and Tables}

\begin{figure*}[h]
    \centering
    \includegraphics[width=0.77\linewidth]{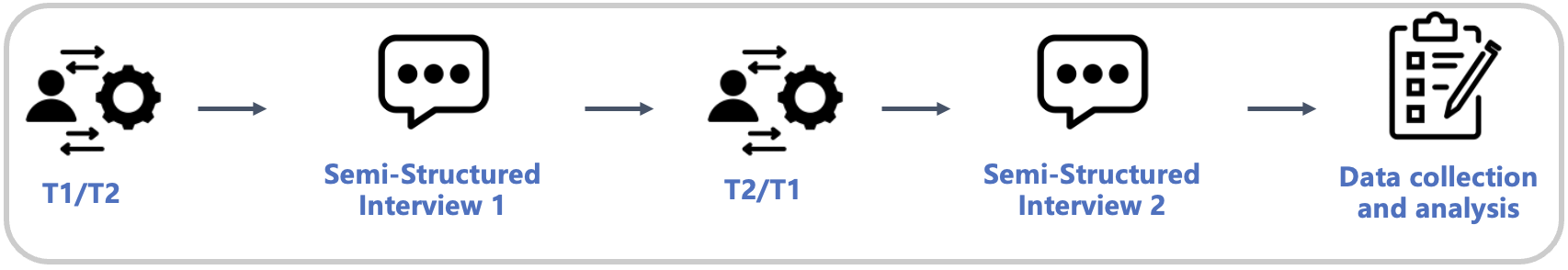}
    \caption{The study flow shows each participant conditioned two tasks (T1/T2) under the supervision and reminder of the investigator.}
    \label{fig:studyflow}
\end{figure*}

\begin{table*}[h]
  \scriptsize
  \centering
  \scriptsize\caption{Summary of democratized information of participants in the user study. Among the 22 participants, 10 were male and 12 were female, aged from 22 to 32. Programming experience is categorized into five levels, from low to high: Level 1 - Little experience; Level 2 - Some experience; Level 3 - Moderate experience; Level 4 - Substantial experience; Level 5 - Professional. P5.js or Processing experience is categorized into four levels, from low to high: Level 1 - Little experience; Level 2 - Some experience; Level 3 - Moderate experience; Level 4 - Substantial experience.}
  \begin{tabular}{p{0.5cm}ccp{3.8cm}p{1.5cm}p{1.5cm}p{2.2cm}p{1.5cm}}
    \toprule
    \textbf{NO.} & \textbf{Gender} & \textbf{Age} & \textbf{Profession} & \textbf{Programming Experience} & \textbf{Programming Language} & \textbf{Other Creative Programming Tool} & \textbf{Used P5.js/Processing} \\
    \hline
    P16 & Male & 26 & VR Artist & 5 & C++, Java, Python, Matlab &  None & 4 \\ \hline
    P19 & Female & 28 & Computer Artist \& Robot Arm Artist & 4 & Java & TouchDesigner, Arduino, Grasshopper & 4 \\
    P20 & Female & 24 & Animation Artist & 4 & Java & None & 3 \\ \hline
    P1 & Female & 24 & Visual Communication Designer  & 3 & Java & Arduino & 3 \\
    P2 & Male & 32 & Computer Artist \& Generative AI Artist & 3 & HTML& TouchDesigner, maxmsp & 3 \\
    P8 & Female & 23 & Industrial Designer & 3 & C\# & None & 3 \\
    P9 & Male & 26 & Computer Artist & 3 & Java & Others & 3 \\
    P11 & Male & 22 & Film Artist & 3 & C\#, Python & None & 1 \\
    P14 & Male & 27 & Computational Creative Artist & 3 & C\#, Java, Python &  N/A & 3 \\
    P17 & Male & 28 & Architect & 3 & Java, Python & Grasshopper & 3 \\
\hline
    P3 & Male & 26 & Environmental Artist \& Computational Creative Artist & 2 & C\# & Arduino, Grasshopper, Houdini & 2\\
    P7 & Female & 22 & Interaction Designer & 2 & C, Python & None & 2 \\
    P10 & Female & 24 & Industrial Designer & 2 & C\# & Arduino & 1 \\
    P12 & Female & 22 & Digital Media Artist & 2 & Java &  None & 3 \\
    P15 & Male & 27 & Interactive Image Artist & 2 & Java &  TouchDesigner, Arduino & 2 \\
    P18 & Male & 22 & Interaction Designer & 2 & C\# & Others & 1 \\
\hline
    P4 & Female & 24 & Product Designer & 1 & Java & Others & 1 \\
    P5 & Female & 26 & Architect & 1 & Python & Arduino & 1 \\
    P6 & Female & 24 & Visual Communication Designer & 1 & N/A & Others & 1 \\
    P13 & Male & 25 & Landscape Artist & 1 & C\# & Others & 1 \\
    P21 & Female & 29 & Industrial Designer & 1 & Python & Arduino & 1 \\
    P22 & Female & 31 & Environmental Artist & 1 & C\# & Others & 1 \\
    \bottomrule
\scriptsize\footnote{different proficiency levels (reported as greater than "Moderate," including "Moderate"  (N=11), "Some," or "Little" (N=11) ) with different programming languages and tools}
  \end{tabular}
  \label{tab:participants}
\end{table*}

\begin{figure*}[h]
    \centering
    \includegraphics[width=0.98\linewidth]{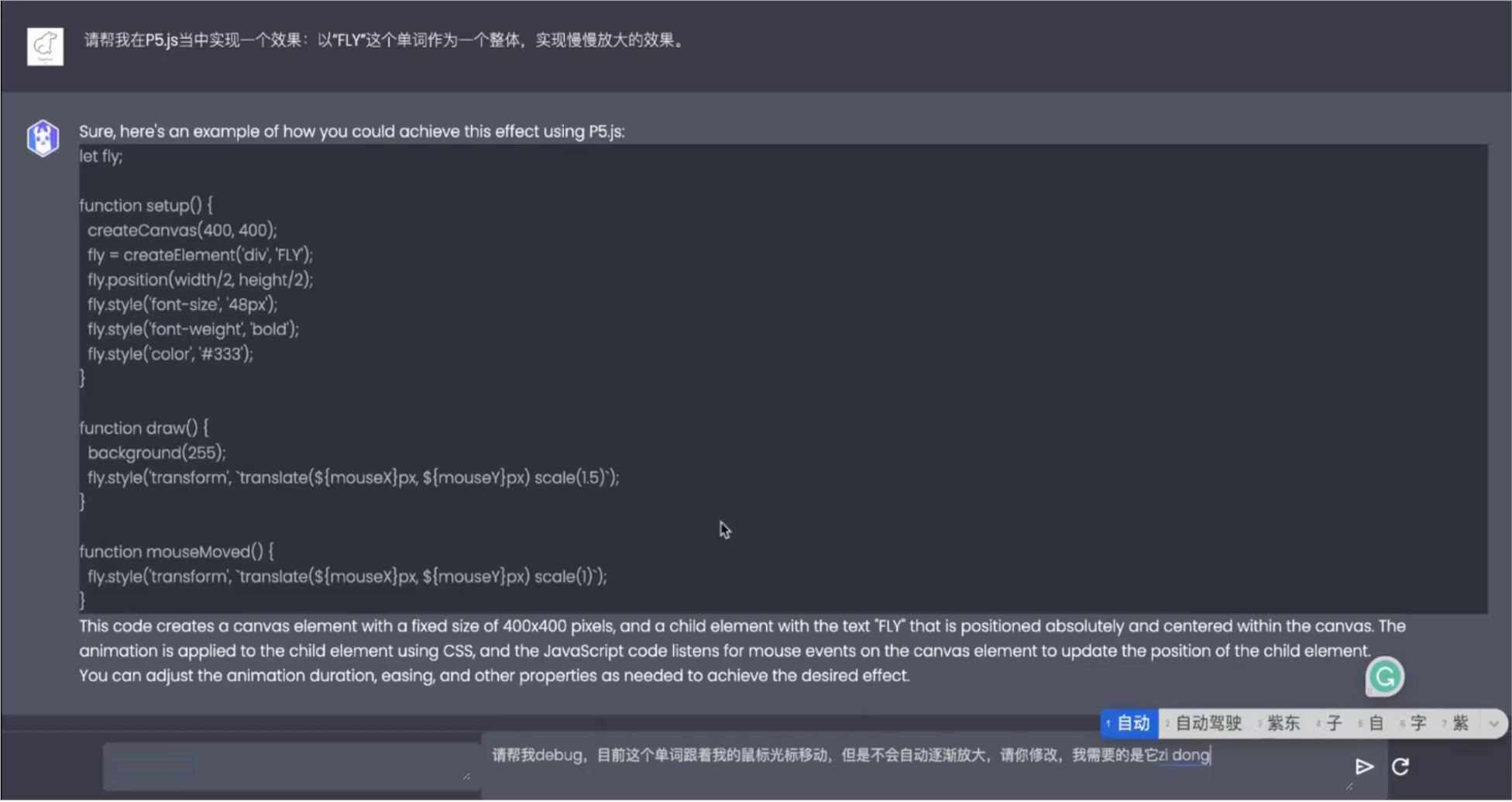}
    \caption{The user interface for Code-Llama we build, based on the layout of ChatGPT, supporting both English and Chinese.}
    \label{fig:screenshot}
\end{figure*}

\section{Semi-Structured Interview Guide}
\label{appendix:interview}

To better understand users' reflection behaviors during creative programming with LLMs, we conducted semi-structured interviews focused on users’ goals, problem-solving strategies, and their reasoning processes during the interaction. The interview guide was designed to elicit rich descriptions of reflection episodes, including triggers, decisions, and outcomes, aligned with the reflection types identified in our analysis (Breakdown, Inquiry, Transformation). Follow-up questions were adapted depending on participants’ responses.

\begin{itemize}
\small
    \item \textbf{General Context and Goals}
    \begin{itemize}
        \item Could you describe what you were trying to co-create with the LLM?
        \item Did you have a clear idea at the beginning, or did it evolve over time?
        \item What were your main goals or intentions behind the project?
    \end{itemize}

    \item \textbf{Interaction with the LLM}
    \begin{itemize}
        \item How did you communicate your ideas to the LLM? 
        \item Were there moments when the LLM misunderstood your prompts? What did you do then?
        \item Can you recall a specific moment when the LLM helped you solve a problem?
    \end{itemize}

    \item \textbf{Breakdown-Oriented Reflection}
    \begin{itemize}
        \item Were there any frustrating or confusing moments during the process?
        \item What did you do when the LLM gave you an unexpected or incorrect output?
        \item Did you ever copy and paste error messages into the LLM? Why or why not?
        \item How did you decide whether to revise the prompt or debug the code yourself?
    \end{itemize}

    \item \textbf{Inquiry-Oriented Reflection}
    \begin{itemize}
        \item Did you try to improve or refine the LLM’s output? How?
        \item Can you describe any specific strategies you used to explore different options?
        \item Did you draw on past experience or knowledge during the process?
        \item How did you evaluate whether an output was “good enough”?
    \end{itemize}

    \item \textbf{Transformation-Oriented Reflection}
    \begin{itemize}
        \item Were there moments when you changed your original plan or goal?
        \item What prompted this change—was it something the LLM did, or something else?
        \item How did you come up with a new direction or idea?
        \item Did you seek outside inspiration, such as code examples or visual references?
    \end{itemize}

    \item \textbf{Wrap-Up}
    \begin{itemize}
        \item Looking back, what was the most valuable thing you learned during the process? 
        \item Is there anything else you’d like to share about your experience?
    \end{itemize}
\end{itemize}

Interview duration: 30–45 minutes.  
All interviews were recorded, transcribed, and analyzed alongside interaction logs and user annotations.

\end{document}